\documentclass[twocolumn,preprintnumbers,amsmath,amssymb,prb]{revtex4}

\usepackage{graphicx}
\usepackage{dcolumn}
\usepackage{bm}

\begin{document}


\title{Sub-MHz Linewidth at 240 GHz from an Injection-Locked Free-Electron Laser}

\author{Susumu Takahashi}
 \email{susumu@iqcd.ucsb.edu}
\author{Gerald Ramian}
\author{Mark S. Sherwin}
\affiliation{Department of Physics and Institute for Quantum and Complex Dynamics, University of California, Santa Barbara CA 93106}%

\author{Louis-Claude Brunel}
\author{Johan van Tol}
\affiliation{National High Magnetic Field Laboratory, Florida State University, Tallahassee FL  32310}%


\date{\today}

\begin{abstract}
Radiation from an ultra-stable 240 GHz solid-state source has been
injected, through an isolator, into the cavity of the University
of California Santa Barbara (UCSB) MM-wave free-electron laser
(FEL). High-power FEL emission, normally distributed among many of
the cavity's longitudinal modes, is concentrated into the single
mode to which the solid state source has been tuned. The linewidth
of the FEL emission is 0.5 MHz, consistent with the Fourier
transform limit for the 2 microsecond pulses.  This demonstration
of frequency-stable, ultra-narrow-band FEL emission is a critical
milestone on the road to FEL-based pulsed electron paramagnetic
resonance spectroscopy.
\end{abstract}

\maketitle

Scientific and technological opportunities in the electromagnetic
spectrum between 0.1 and 10 terahertz (THz) have fueled rapid
growth in the technology for generating THz radiation.~\cite{doe,
siegel02} THz sources can now generate coherent broad-band
radiation suitable for spectroscopic~\cite{siegel02} and
high-field nonlinear~\cite{you93, carr02} experiments, as well as
narrow-linewidth radiation at fixed~\cite{siegel02, kohler02} or
tunable~\cite{vdi} frequencies. Free-electron lasers (FELs) are
proven sources of tunable high-power THz
radiation.~\cite{ucsbfel,felix} Almost all FELs are driven by
radio-frequency linear accelerators (RF-LINACs), and emit pulses
with a typical duration measured in picoseconds. Such pulses have
relatively broad, Fourier-transform limited linewidths. The
University of California Santa Barbara (UCSB) FELs are driven by
an electrostatic accelerator. Under free-running operation, these
FELs generate pulses which are several microseconds long but lase
simultaneously on a number of modes.

An important niche has remained unfilled: tunable, high-power ($>$
100 W) THz pulses with intrinsically narrow linewidths. Such
sources are desirable for nonlinear experiments involving very
narrow excitation lines, such as pulsed electron paramagnetic
resonance (EPR), as well as nonlinear spectroscopy of dilute
molecular gases and Rydberg atoms.

This letter shows that the frequency of UCSB's MM-wave FEL
(120-890 GHz) can be locked to the frequency of an ultra-stable
solid-state oscillator. The experiment demonstrates that the
important niche for tunable, high-power, THz sources with
intrinsically narrow linewidths can be filled using
electrostatic-accelerator driven FELs. It also represents a key
milestone on the road to high-power pulsed EPR at frequencies of
240 GHz and above.

\begin{figure}
\includegraphics[width=70 mm]{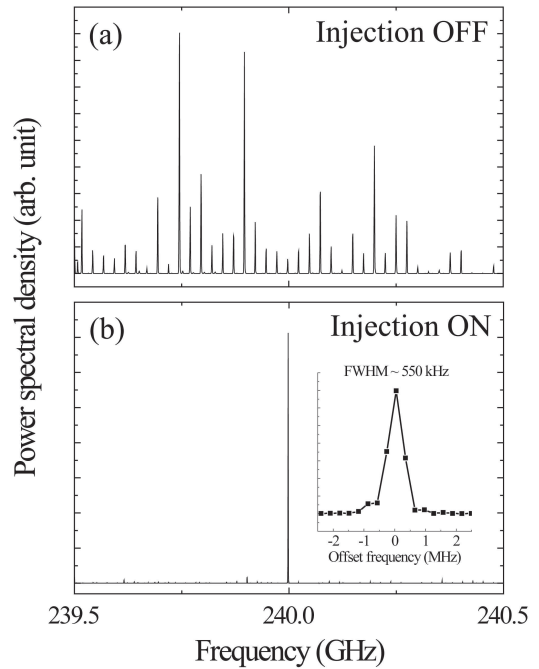}
\caption{\label{fig:data} FEL spectra. (a)FEL spectrum without the
injection-locking. The spectrum shows many longitudinal modes
spaced by 25 MHz. (b)FEL spectrum with injection source on.
Single-mode lasing is observed with no pulse-to-pulse frequency
fluctuations.}
\end{figure}

\begin{figure*}
\includegraphics[width=140 mm]{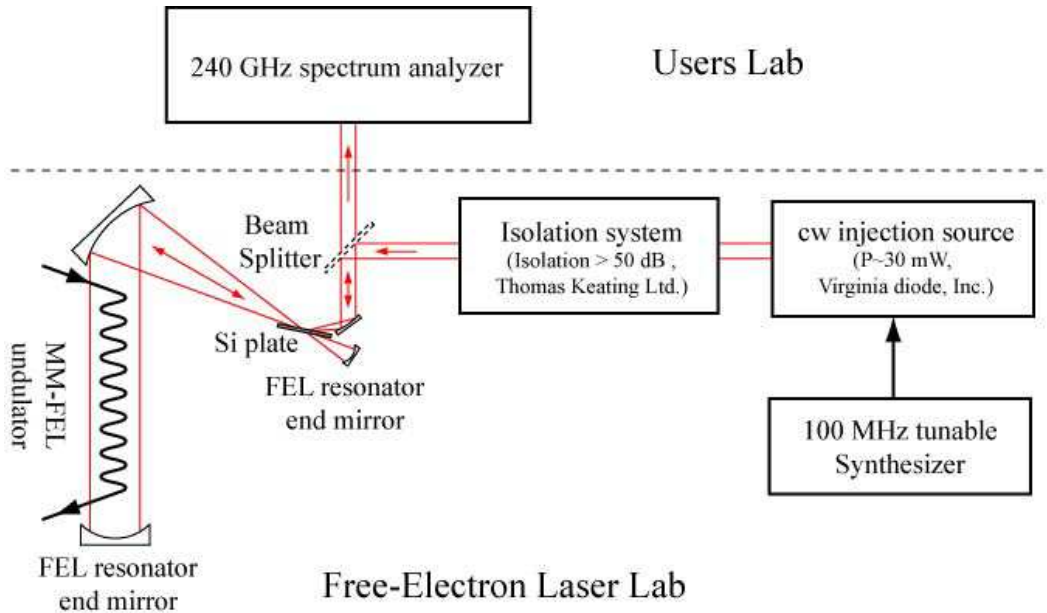}
\caption{\label{fig:injection} Schematic of the injection-locking
system for the UCSB MM-FEL. The injection source, isolator and
tunable 100 MHz synthesizer are located in free-electron laser
lab. The FEL outputs are sent to the users lab through a vacuum
optical transport system.}
\end{figure*}

The spectrum of a single pulse of the free-running FEL is shown in
Fig.~1(a). The spectrum shows many modes with the 25 MHz spacing
which correspond to longitudinal modes of the 6 m long FEL cavity.
The FEL spectrum and center frequency are slightly different on
each pulse (pulse-to-pulse fluctuation of center frequency $\sim
0.05 \%$). Pulses with spectra like these are routinely used for
many experiments. However, pulsed EPR requires a much narrower
spectrum.

A well-known technique to force a laser to emit on a single mode
is to inject into the laser cavity a weak seed beam from an
oscillator with a very stable and well-defined frequency. This
technique is called injection-locking. If the power of the seed
beam is significantly larger than the power in the random
electromagnetic fluctuations that are usually amplified for
lasing, then only the seed beam will be amplified. Fig.~1(b) is
the spectrum with injection on. The effect is immediate and
unequivocal. Lasing always occurs at one frequency on the same
longitudinal mode selected by the injection source frequency. The
injection-locking is observed only when the source frequency
matches one of the resonator's longitudinal modes. The full-width
half maximum (FWHM) width of the lasing line is $\sim$ 550 kHz
which is close to the Fourier transform limit of the sampling
window ($\sim 2 \mu s$).

Fig.~2 illustrates the UCSB injection-locking system. The cw
injection source, built by Virginia Diode Inc.(VDI),~\cite{vdi}
generates as much as 30 mW at 240 GHz. The VDI source consists of
a 15 GHz phase-locked oscillator, RF active frequency doubler, and
three varactor doubler stages. The source frequency is tunable
within small frequency range by employing a tunable 100 MHz
digital synthesizer as a reference. This tunability is essential
to match the injection source frequency to the FEL resonator and
to facilitate operation of the injection-locking system. A key
component is a quasi-optical Faraday-rotation, isolator developed
by Thomas-Keating Ltd.~\cite{tk} that provides $>$ 50 dB
isolation. Insertion loss is about 6 dB. The isolator is essential
to protect the injection source from the FEL power after startup.
For this injection-locking system, the FEL is run in a two arm
configuration as shown in Fig.~2. A silicon plate, offset slightly
from Brewster's angle, couples injection power in and FEL power
out. This allows full coupling to the FEL resonator's fundamental
mode. The Si plate and the FEL end mirrors are tuned to optimize
FEL performance. A second beam-splitter directs about $90 \%$ to
the users lab. A spectrum analyzer located in the users lab
analyzes the FEL output.

\begin{figure}
\includegraphics[width=70 mm]{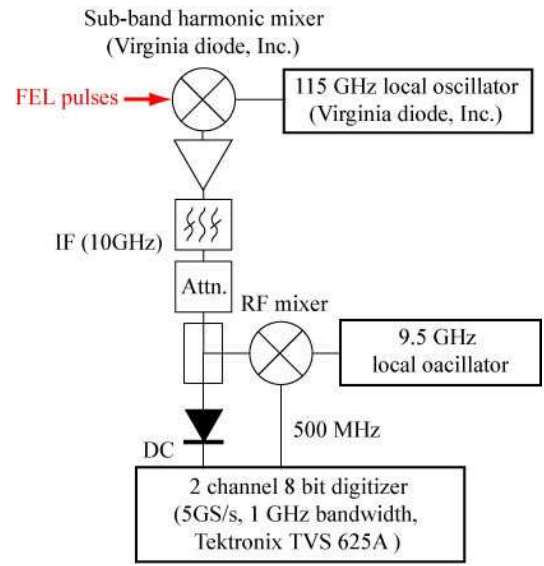}
\caption{\label{fig:sa} Circuit diagram of the 240 GHz spectrum
analyzer.}
\end{figure}

The spectrum analyzer is based on a heterodyne detection system.
240 GHz FEL radiation is attenuated and directed to a
subharmonically-pumped mixer (VDI) driven at 115 GHz by a second
VDI multiplier chain as Local Oscillator. A 10 GHz intermediate
frequency (IF) signal is then amplified, filtered, and sent to a
second mixer to further downconvert to 500 MHz. A bandpass filter
in the IF chain is used to remove negative frequency artifacts. A
2 channel, 8 bit, high performance digitizer (Tektronix TVS 625A)
measures both signals and passes the digitized signals to a
computer for fast Fourier transform processing. The sampling rate
of the digitizer is 5 GS/s and the bandwidth is 1 GHz. The whole
spectrum analyzer is controlled by LabView programs which record
the single pulse FEL spectra. The IF signal is also detected by a
10 GHz square-law detector.

It is instructive to compare this injection-locking demonstration
with a previous experiment in 1991.~\cite{amir91} That experiment
used a molecular gas laser which is not tunable as an injection
source at 2.5 THz for the UCSB FIR FEL (890 GHz - 4.7 THz). The
FEL cavity length had to be adjusted to match the longitudinal
mode frequency to the injection source. Evidence for
injection-locking was compelling but indirect, as the power
spectrum of the emission could not be measured. The tremendous
improvements in tunable solid state oscillators in the intervening
16 years have made injection-locked MM-wave FEL at UCSB a robust
tool for experimentation.

We are developing a high frequency pulsed EPR spectrometer to
investigate the structure and dynamics of biological molecules in
aqueous solution. This setup includes the injection-locked FEL,
running at 240 GHz, a 9 T superconducting magnet with high
homogeneity, a quasi-optical pulse generator~\cite{hegmann00,
doty04} and quasi-optical delay line~\cite{allen07}. Details of
the EPR setup will be described elsewhere.

Currently, most high-power pulsed EPR spectrometers are near the
X-band frequency of 9.5 GHz with kW-level power. A trend in the
evolution of next generation pulsed EPR is for higher frequency
both for a finer spectral resolution and a better resolution for
fast structural changes due to less motional
averaging.~\cite{freed00} Currently there exist only a few
high-power pulsed EPR spectrometers at frequencies above 50 GHz. A
1 kW Klystron amplifier-based 95 GHz system operates at
Cornell.~\cite{hofbauer04} Another Klystron based-pulsed EPR
system is being built at St. Andrews UK. A Gyrotron-amplifier may
also be useful for high-power pulsed EPR. Although the systems are
currently not used for pulsed EPR, 10-15 W 250 GHz and 14 W 140
GHz Gyrotron-based system have been built at MIT.~\cite{bajaj03,
joye06} Compared with these other sources, the UCSB FEL has
significant advantages for pulsed EPR at 240 GHz and higher
frequencies. First, the FEL is the most powerful source at 240
GHz. Moreover the UCSB FEL is widely tunable from 120 GHz to 4.7
THz, which covers g=2 EPR signals from 5 T to higher than magnetic
fields produced by any existing DC-field
magnets.~\cite{highestmagnet} For the UCSB MM-FEL, 240 GHz is at
the low end of its frequency range and can produce hundreds of
watts of power. It can produce much higher power at higher
frequencies. Thus the FEL will work even better for pulsed EPR at
a higher frequency.

In summary, we have demonstrated injection-locked FEL operation
with sub-MHz bandwidth using a tunable solid state source as an
injection seed. The high-power, tunability, and extremely narrow
linewidth of the FEL opens up the possibility for new applications
such as high-power, high-frequency, pulsed EPR spectroscopy.

We wish to thank David Enyeart for his support of the
injection-locking system installation and FEL operation. This work
was supported by NSF grants (DMR-0321365 and DMR-0520481).


\begin{thebibliography}{18}
\bibitem[{hig()}]{highestmagnet}
\bibinfo{note}{The world's highest DC field magnet is the 45 tesla hybrid
  magnet in the National High Magnetic Field Laboratory (NHMFL). 45 T
  corresponds to 1.26 THz for g=2 EPR.}
\end{thebibliography}

\end{document}